\documentclass[a4paper]{jpconf}
\usepackage{graphicx}
\usepackage{aas_macros,comment}
\begin{document}
\title{Modelling Fast-Alfv\'en Mode Conversion Using SPARC}

\author{H Moradi, P S Cally}

\address{Monash Centre for Astrophysics, School of Mathematical Sciences, Monash University, Victoria 3800, Australia}

\ead{hamed.moradi@monash.edu}

\begin{abstract}
We successfully utilise the SPARC code to model fast-Alfv\'en mode conversion in the region $c_A \gg c_S$ via 3-D MHD numerical simulations of helioseismic waves within constant inclined magnetic field configurations. This was achieved only after empirically modifying the background density and gravitational stratifications in the upper layers of our computational box, as opposed to imposing a traditional Lorentz Force limiter, to ensure a manageable timestep. We found that the latter approach inhibits the fast-Alfv\'en mode conversion process by severely damping the magnetic flux above the surface. 
\end{abstract}

\section{Introduction}
A series of recent studies (e.g., \cite{cg2008,ca2010,ch2011,hc2011,kc2011,kc2012,felipe2012,hc2012}) have shown that an important feature not widely accounted for in sunspot seismology is fast-Alfv\'en wave mode conversion - a process which occurs at, and beyond, the fast wave reflection height (where Ê$c_A Ê\approx Ê\omega/k_h$; Ê$c_A$ denotes the Alfv\'en wave speed, $\omega$ the wave frequency and $k_h$ the horizontal wavenumber) -  being spread over many scale heights for wavenumbers typical of local helioseismology. This process is most efficient for $\theta$ (field inclination from vertical) between $30^\circ-40^\circ$, and $\phi$ (angle between the magnetic field and wave propagation planes) between $60^\circ- 80^\circ$, and appears to have the potential to modify the seismic wave-path through the solar atmosphere, thereby affecting the wave travel times that are the basis of our inferences about the subsurface (see \cite{gizonetal2009,moradietal2010,gbs2010,moradi2012} for recent reviews). Motivated by these studies, our aim is to use the Seismic Propagation through Active Regions and Convection (SPARC) code, a 3-D magnetohydrodynamic (MHD) wave-propagation code developed by \cite{hanasoge2007} for computational heliosiesmology, to numerically simulate this process, and investigate the implications of fast-Alfv\'en wave mode conversion on the seismology of the photosphere. In this paper we describe our attempts at forward modelling this process with SPARC using a quiet-Sun background model permeated by homogenous inclined magnetic fields. 

\section{Numerical Setup}

The SPARC code solves the 3-D linearized Euler and induction equations of magnetofluid motion in Cartesian geometries to investigate wave interactions with local perturbations (e.g., sound speed, pressure, density, flows, magnetic field etc.). Over the past few years, a number of various solar phenomena have been studied using this code (e.g., \cite{hdc2007,hanasoge2008,birchetal2009,mhc2009,hdd2010}). The computational box we employ for our simulations using SPARC spans $186.6$ Mm in the horizontal direction (128 evenly-spaced grid points in the horizontal directions $x$ and $y$; $\Delta x= \Delta y=1.46$ Mm/pixel), and from $2.5$ Mm above the surface ($z=0$; where $z$ denotes height in Mm) to $25$ Mm below in the vertical direction (300 non-uniformly spaced grid points in $z$; $\Delta z$ varies from several hundred kilometres at depth, to tens of kilometres in the near-surface layers). The vertical boundaries of the box are absorbent, with perfectly matched layer (PML) boundary layers spanning the top 10 and bottom 7 grid points in $z$, while periodic boundary Êconditions are imposed on the horizontal sides. In a similar manner to \cite{kc2011,kc2012,felipe2012}, we use a monochromatic ($\nu = \omega/2\pi=5$ mHz) plane-parallel wave driver, which only excites waves propagating in the ($x$, $z$) planes. We do this by imposing a perturbation of the form:
\begin{equation}
\sin(\omega t)~\e^{-(x-x_0)^2/2\delta_x^2}~\e^{-(z-z_0)^2/2\delta_z^2}
\end{equation}
in a few grid points near $z_0 = -5$ Mm and centred at around $x_0$ = 0 Mm, in pressure, density and velocity ($x$ and $z$ components only). 

\section{Model Atmosphere}

The Êbackground model used is a convectively stabilised solar model Ê(CSM\_B) Êfrom Ê\cite{schunkeretal2011}. On top of this background model we employ a constant, Êinclined magnetic Êfield configuration using the prescription from \cite{cg2008}: Ê	
\begin{equation}
\mathbf{B}_0 Ê= ÊB_0 Ê(\sin\theta \cos\phi, \sin\theta\sin\phi, \cos\phi), 	
\end{equation}
We choose $B_0 = 1500$ G and a number of different angle orientations in the range $0 < \theta < 90^\circ$ and $0 < \phi < 90^\circ$. 

In 3-D MHD simulations, the timestep ($\Delta t \sim  \Delta z/c_A$) is often highly constrained by the Courant-Friedrichs-Lewy (CFL) condition, due to the exponentially increasing value of $c_A=\mathbf{B}_0/(\mu_0\rho)^{1/2}$ (where $\rho$ denotes density and $\mu_0$ the magnetic permeability) above the surface. This causes the wavelengths of both the fast and Alfv\'en waves to become quite large, resulting in an extremely stiff numerical problem. The most common way of dealing with this issue, in computational helioseismology, has been to simply apply a $c_A$ ``limiter'' to moderate the action of the Lorentz Force when the ratio $c_A$/$c_S$ (where $c_S$ denotes sound speed) becomes exceedingly large. Some choices for these limiters have been discussed before (e.g., \cite{cgd2008,cameronetal2011,hanasoge2008,rsk2009,braunetal2012}). Generally, they tend to prefix the Lorentz-Force terms in the momentum equations, with $c_A$ typically being capped at $\sim 20-60$ km/s. The physical implications of artificially limiting the Lorentz Force in such a manner (particularly on the seismology) have not been explored. Recently though \cite{cally2012} has shown that significant internal reflection of Alfv\'en waves can occur if the $c_A$ profile is not adequately treated above the surface. 

\begin{figure}[htb]
\begin{minipage}{18pc}
\includegraphics[width=17pc,trim=1cm 5cm 2cm 6cm, clip=true]{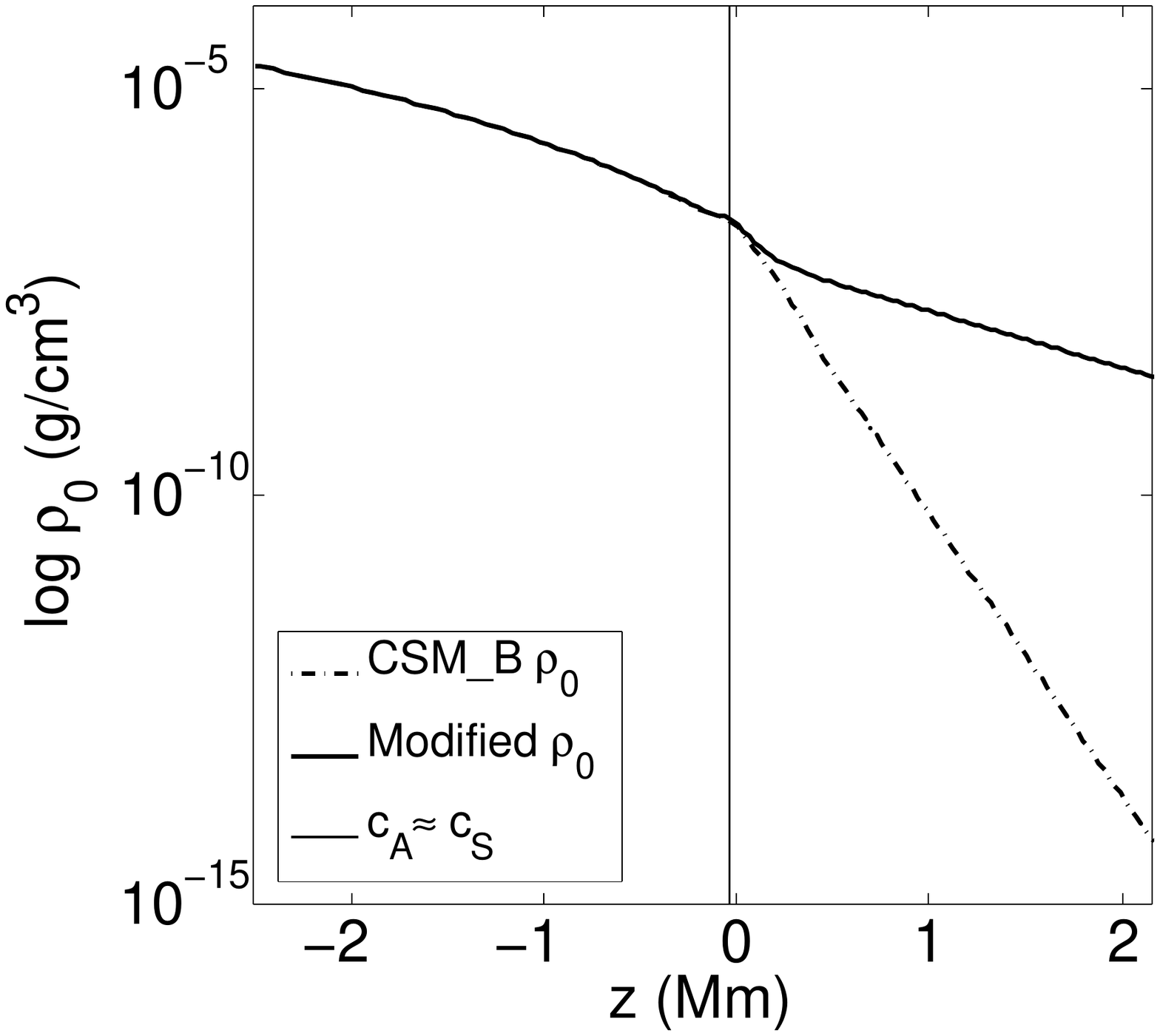}
\end{minipage}\hspace{2pc}%
\begin{minipage}{18pc}
\includegraphics[width=17pc,trim=1cm 5cm 2cm 6cm, clip=true]{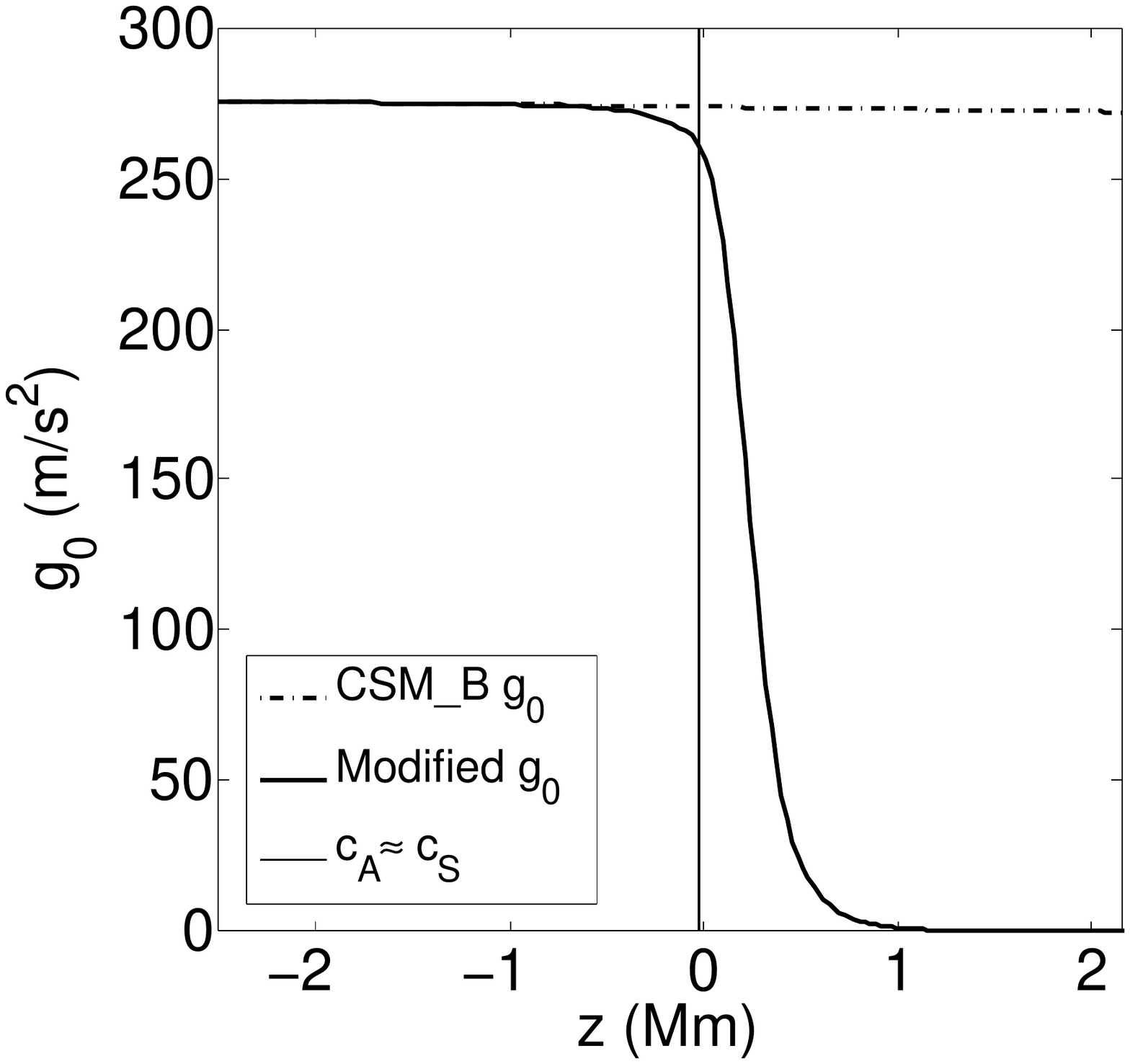}
\end{minipage}
\caption{\label{fig:back} Plots of the original CSM\_B (\cite{schunkeretal2011}; dot-dashed line) and modified (solid line) background density (left) and gravitational acceleration (right) profiles as a function of height. The solid vertical line represents the $c_A \approx c_S$ height.}
\end{figure}

\begin{figure}[htb]
\begin{minipage}{18pc}
\includegraphics[width=17pc,trim=0cm 4cm 2cm 5cm, clip=true]{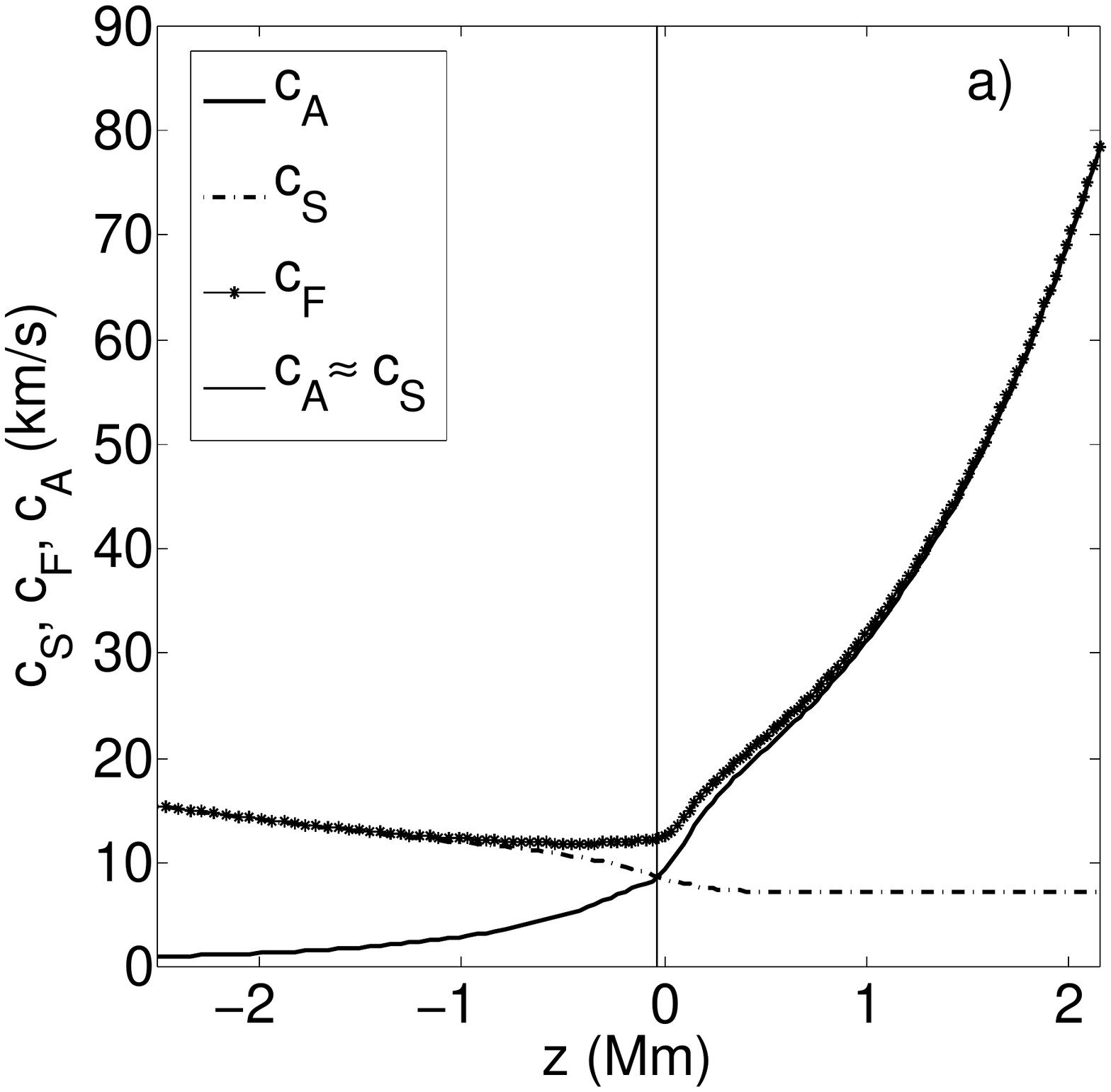}
\end{minipage}\hspace{2pc}
\begin{minipage}{18pc}
\includegraphics[width=17pc,trim=0cm 4cm 2cm 5cm, clip=true]{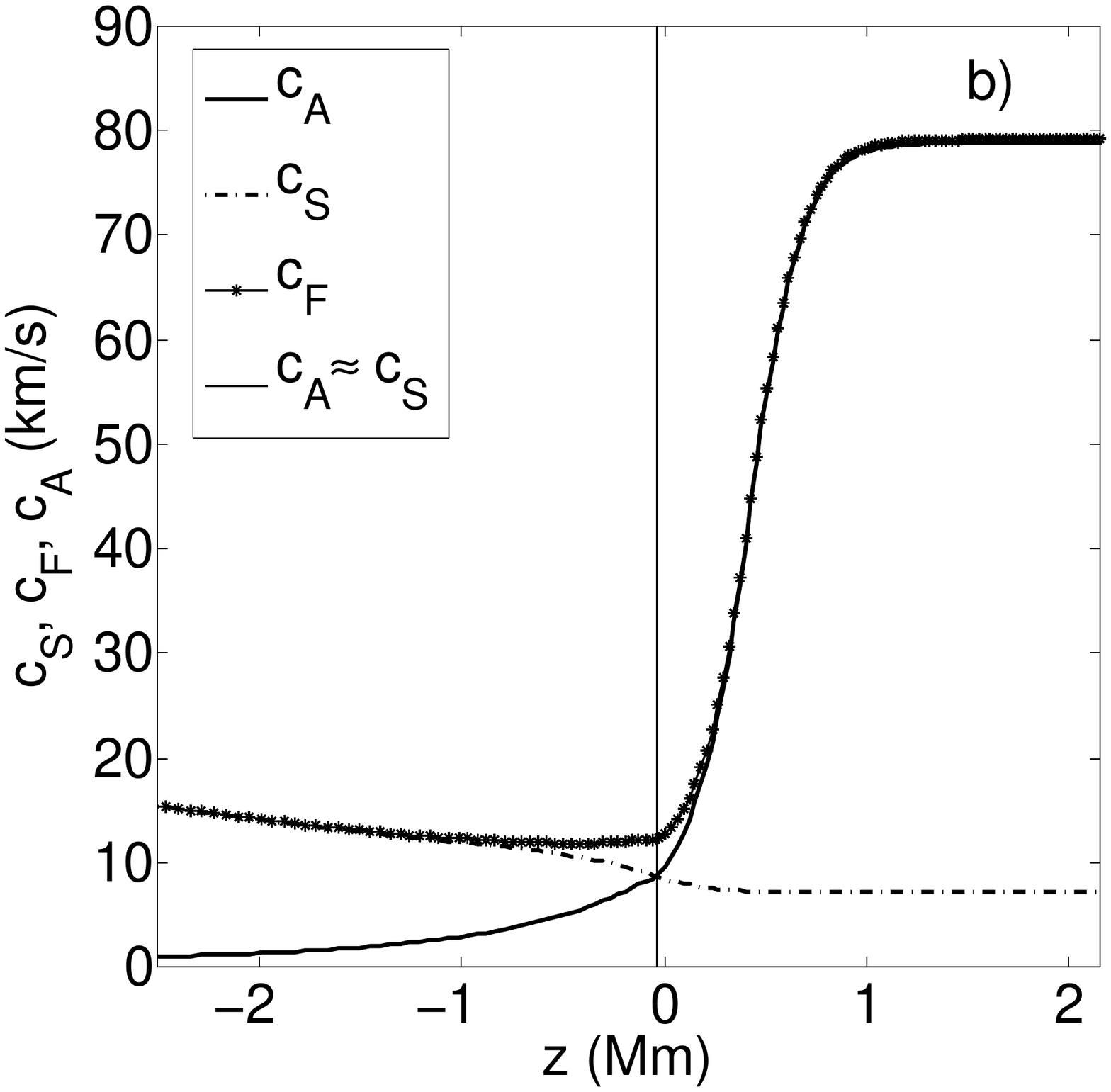}
\end{minipage}  
\caption{\label{fig:speeds}a): Plots of various wave speeds as a function of height resulting from the modified CSM\_B background model. b): Same as a) but for an unmodified CSM\_B background in conjunction with a Lorentz Force limiter (i.e, the Lorentz Force is multiplied by a factor of $200c_S^2/(200c_S^2 + c_A^2)$, leading to $c_A$ and $c_F$ being capped at $\approx 80$ km/s).  The solid vertical line represents the $c_A \approx c_S$ height.}
\end{figure}

Our method of ensuring a reasonable $\Delta t$ for our calculations involves empirically modifying the background density ($\rho_0(z)$) in the upper layers ($\sim 0.5<z<2.5$ Mm) of CSM\_B, in conjunction with a commensurate modification of the gravitational Êacceleration ($g_0(z)$) profile over the same $z$ range (in Êorder Êto Êoffset Êthe Êchange Êin Êthe Ê density Êscale Êheight), to obtain a maximum global fast ($c_F$) Êand Ê$c_A$ Êof Ê$\approx Ê80$ Êkm/s. All other background variables remain unaltered. The modified profiles of $\rho_0(z)$ and $g_0(z)$ are shown in Figure~\ref{fig:back}. The value of $80$ km/s was chosen because it provides us with a reasonable time step ($\Delta t =0.5$ s), and is safely higher than the largest horizontal phase-speeds Ê($\omega/k_h$) Êtypically sampled in sunspot Êseismology (e.g., \cite{couvidatetal2005}). This is important since $\omega/k_h\approx c_A$ also denotes the location of the fast mode reflection height in the solar atmosphere \cite{kc2012}. Figure~\ref{fig:speeds} a) shows the resulting $c_A$ and $c_F$ profiles in our model as a function of height. For comparison purposes, the $c_A$ and $c_F$ profiles which would result from imposing a Lorentz Force limiter, instead of modifying $\rho_0(z)$ and $g_0(z)$ above the surface, are shown in Figure~\ref{fig:speeds} b).

The obvious downside of empirically modifying $\rho_0(z)$ and $g_0(z)$ in order to satisfy the CFL condition is that background model will no longer be as `solar-like' (i.e., in terms of eigenfrequencies, eigenfunctions and power spectrum) as CSM\_B. The larger $\rho_0(z)$ profile which now results above the surface also modifies acoustic cut-off frequency ($\omega_{ac} = c_S/2H_{\rho}$; where $H_\rho$ denotes the density scale height), which is reduced from $\nu =  5 .2$ mHz to $3.6$ mHz. We also find that the modified atmosphere produces large-amplitude convective ($g$-) modes Êat Ê$\nu \approx 1.5 - 1.7$ mHz (it is worth noting though that this is a frequency range which is typically associated with supergrannulation noise and is generally filtered out/ignored in sunspot seismology). 

However, more importantly for our concerns, this method ensures that we satisfy our CFL condition without any direct Êmodification of the Lorentz Force via the introduction of an artificial term in Maxwell's equations, which as we shall show in the proceeding section, results in unphysical damping of the magnetic flux Êabove Êthe surface and inhibits the fast-Alfv\'en mode conversion process. 	

\section{Results}
Following \cite{kc2011,kc2012,felipe2012}, we use velocity projections onto three orthogonal directions ($\mathbf{\hat e}_{long}$, which selects the longitudinal component of wave propagation, i.e, the slow mode; $\mathbf{\hat e}_{trans}$, which selects the transversal component of wave propagation, i.e., the fast mode; $\mathbf{\hat e}_{perp}$, which selects the perpendicular component of wave motion, i.e., the Alfv\'en mode; see equations 1-3 in \cite{kc2011} for definitions) to separate the Alfv\'en mode from the fast and slow magneto-acoustic modes in the region $c_A\gg c_S$. We also calculate the temporally Êaveraged Êacoustic Ê($\mathbf{F}_{ac} Ê= Ê\langle p_1\mathbf{v}_1 \rangle$; where $p$ represents pressure and $\mathbf{v}$ represents the 3-D velocity respectively, subscript ``1'' represents perturbations), magnetic ($\mathbf{F}_{mag} Ê= Ê\langle \mathbf{B}_1 Ê\times Ê(\mathbf{v}_1 Ê\timesÊ\mathbf{B}_0)/\mu_0 \rangle$) and total ($\mathbf{F}_{tot} = \mathbf{F}_{ac}Ê+ \mathbf{F}_{mag}$) energy fluxes in order to measure the efficiency of conversion to Alfv\'en waves around the $c_A \approx c_S$ equipartition height. 
\begin{figure}[ht]
\begin{minipage}{14pc}
\includegraphics[width=13.5pc,trim=0cm 1cm 1cm 1cm, clip=true]{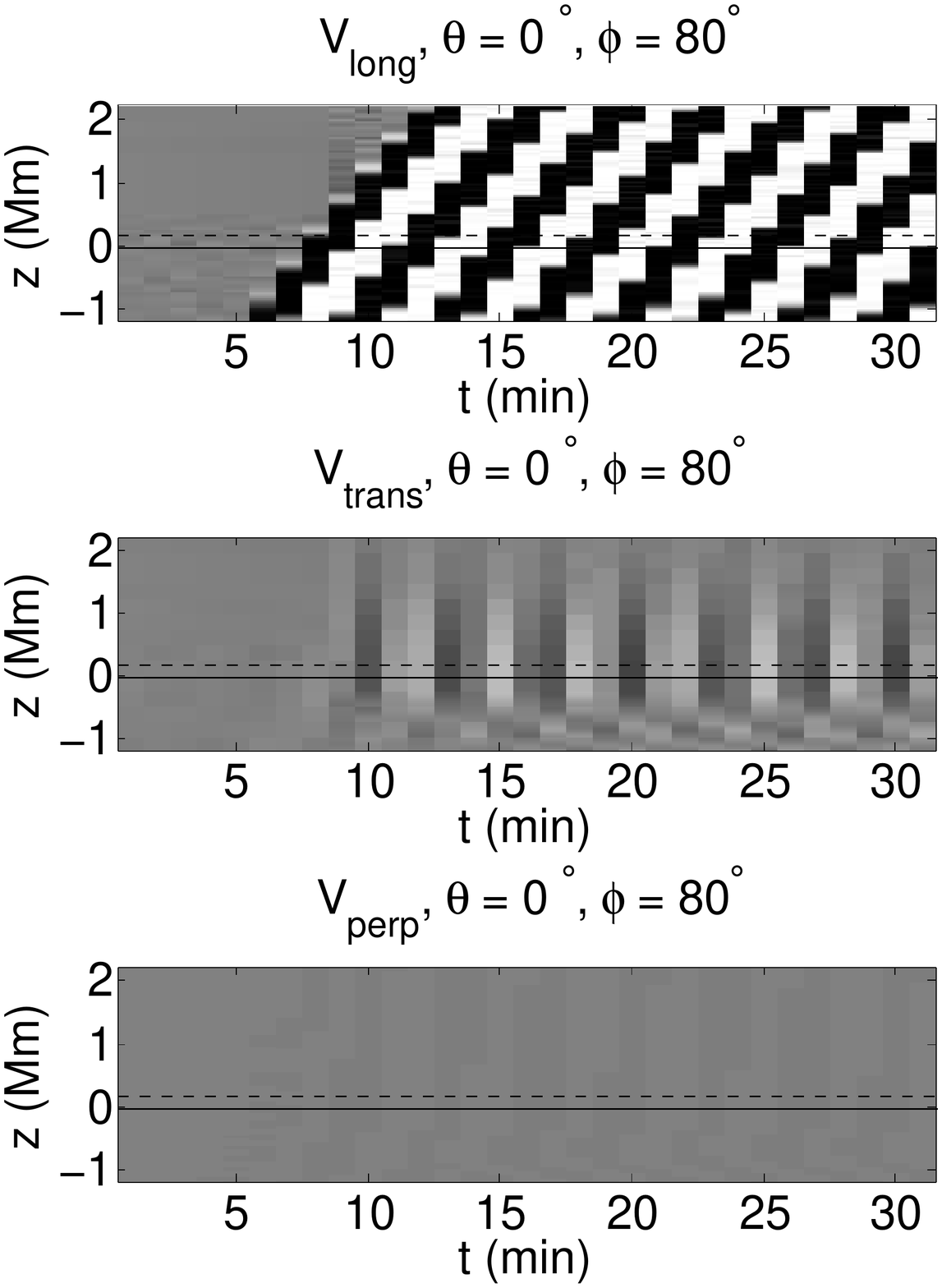}
\end{minipage}\hspace{-1.5pc}%
\begin{minipage}{14pc}
\includegraphics[width=13.5pc,trim=0cm 1cm 1cm 1cm, clip=true]{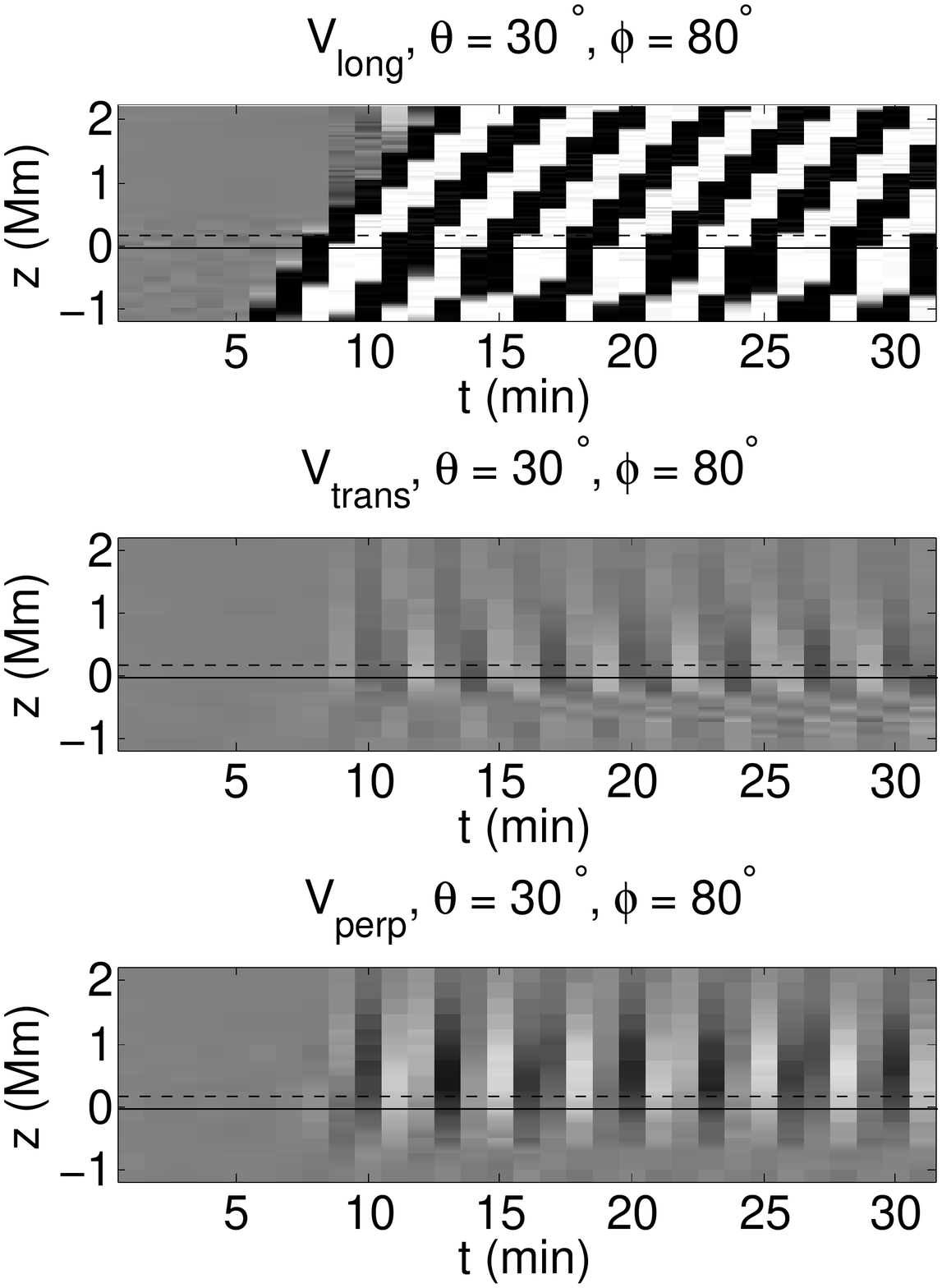}
\end{minipage}\hspace{-1.5pc}%
\begin{minipage}{14pc}
\includegraphics[width=13.5pc,trim=0cm 1cm 1cm 1cm, clip=true]{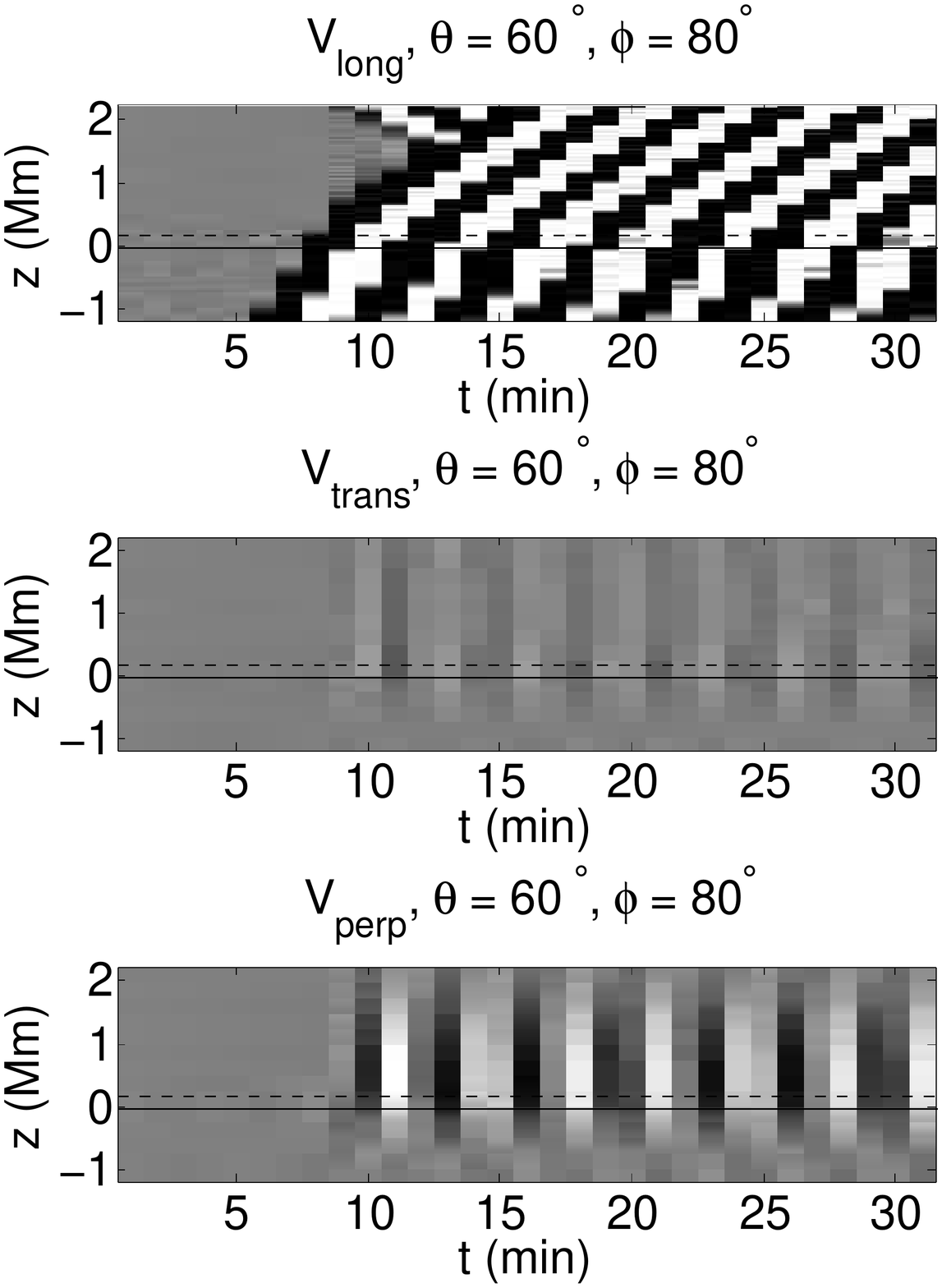}
\end{minipage}  
\caption{\label{fig:proj} Figures above represent the velocity projections derived from simulations using constant inclined magnetic field configurations~(left column: $\theta=0^{\circ}, \phiÊ=Ê80^{\circ}$; middle column: $\theta= 30^{\circ}, \phi=80^{\circ}$; right column: $\theta= 60^{\circ}, \phi=80^{\circ}$) as a function of height (Mm) and Êtime (minutes). $V_{perp}$ andÊ $V_{trans}$ amplitudes have been scaled by a factor of $(\rho_0 c_A)^{1/2}$, while $V_{long}$ has been scaled by a factor of $(\rho_0c_S)^{1/2}$. The grayscale is the same in all panels. The horizontal solid and dotted lines represent the $c_A Ê\approxÊc_S$ and fast-mode reflection heights respectively.}
\end{figure}

\begin{figure}[ht]
\begin{minipage}{13.5pc}
\includegraphics[width=13pc,trim=0cm 6cm 2cm 6cm, clip=true]{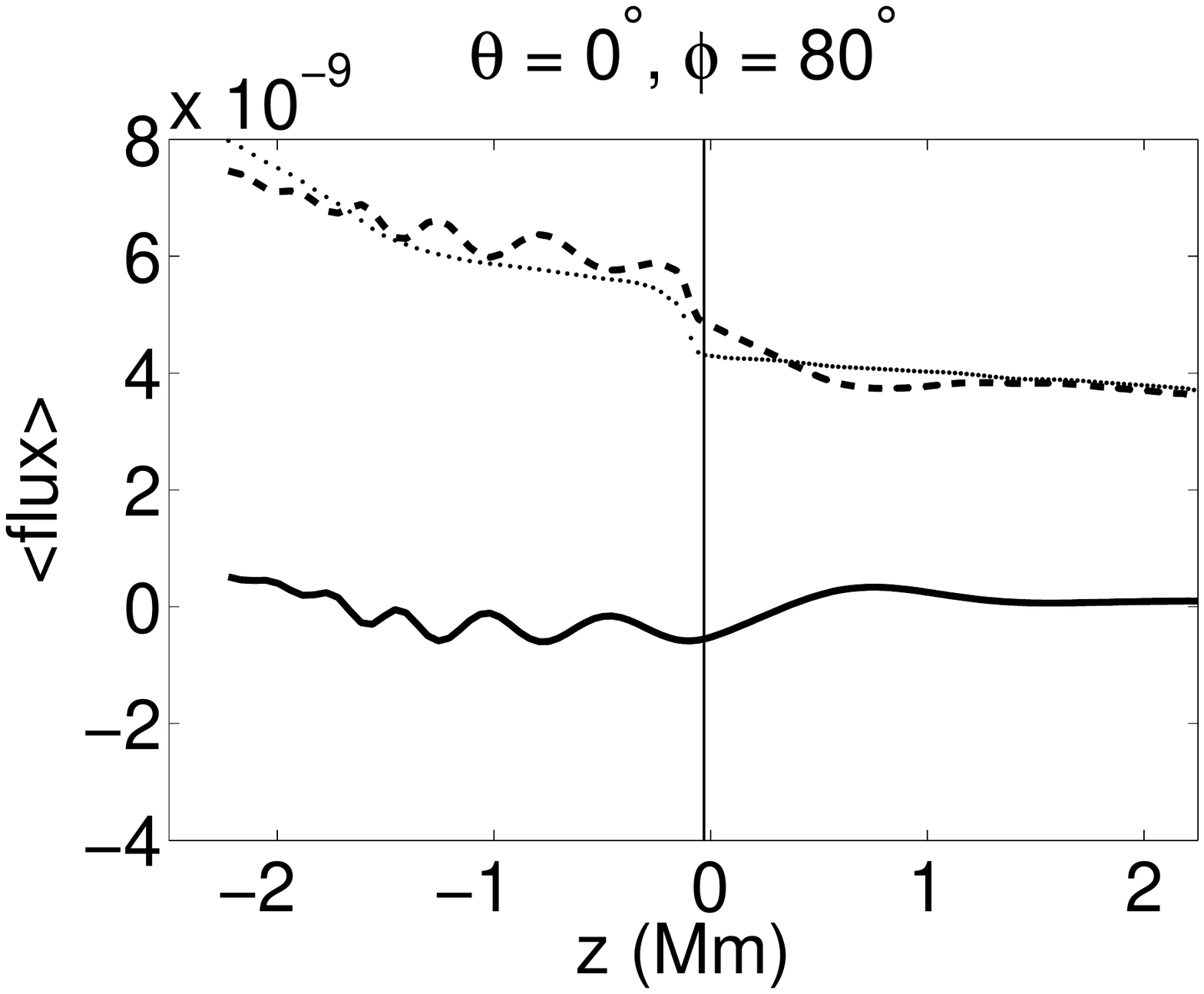}
\end{minipage}\hspace{-1pc}%
\begin{minipage}{13.5pc}
\includegraphics[width=13pc,trim=0cm 6cm 2cm 6cm, clip=true]{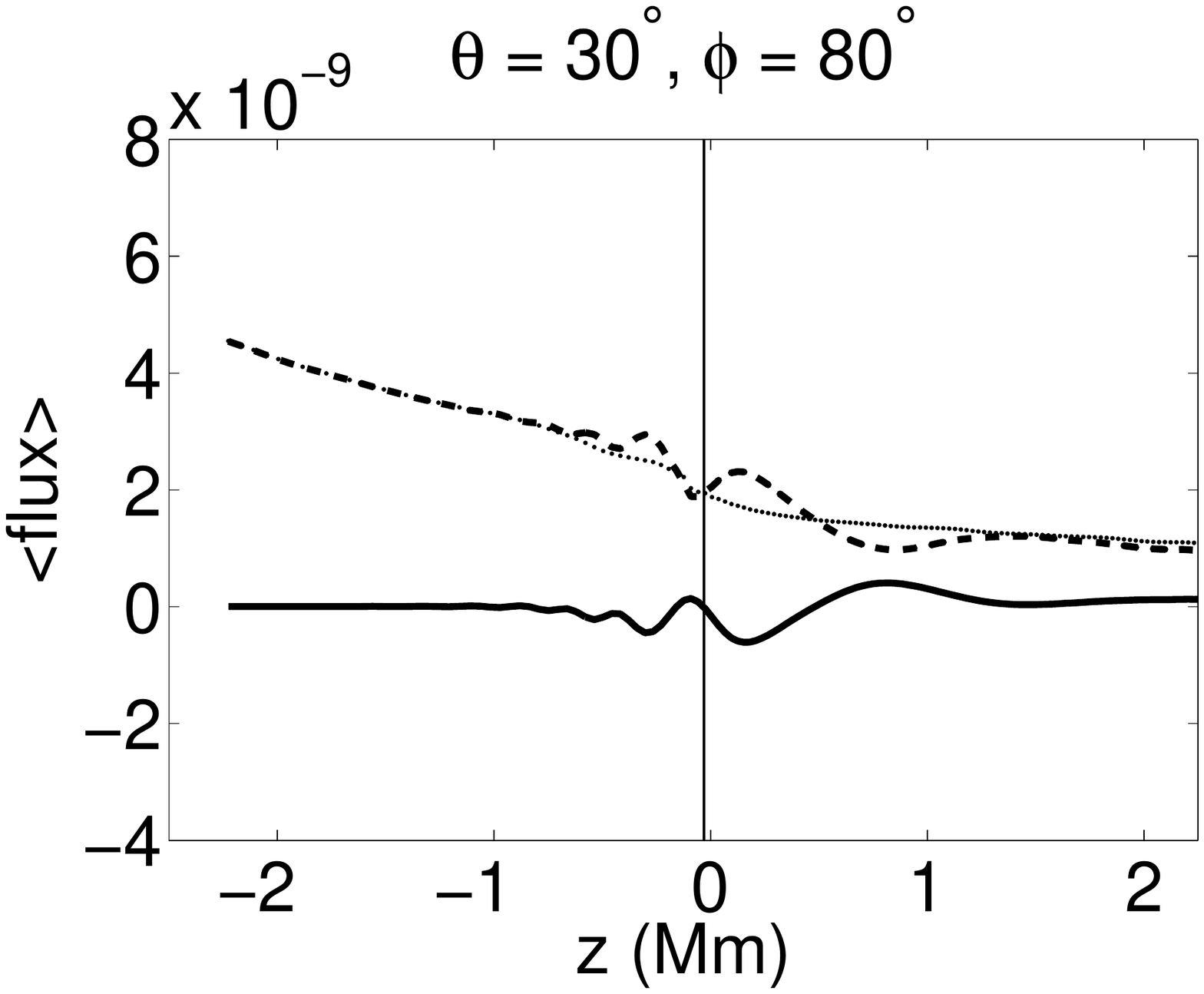}
\end{minipage}\hspace{-1pc}%
\begin{minipage}{13.5pc}
\includegraphics[width=13pc,trim=0cm 6cm 2cm 6cm, clip=true]{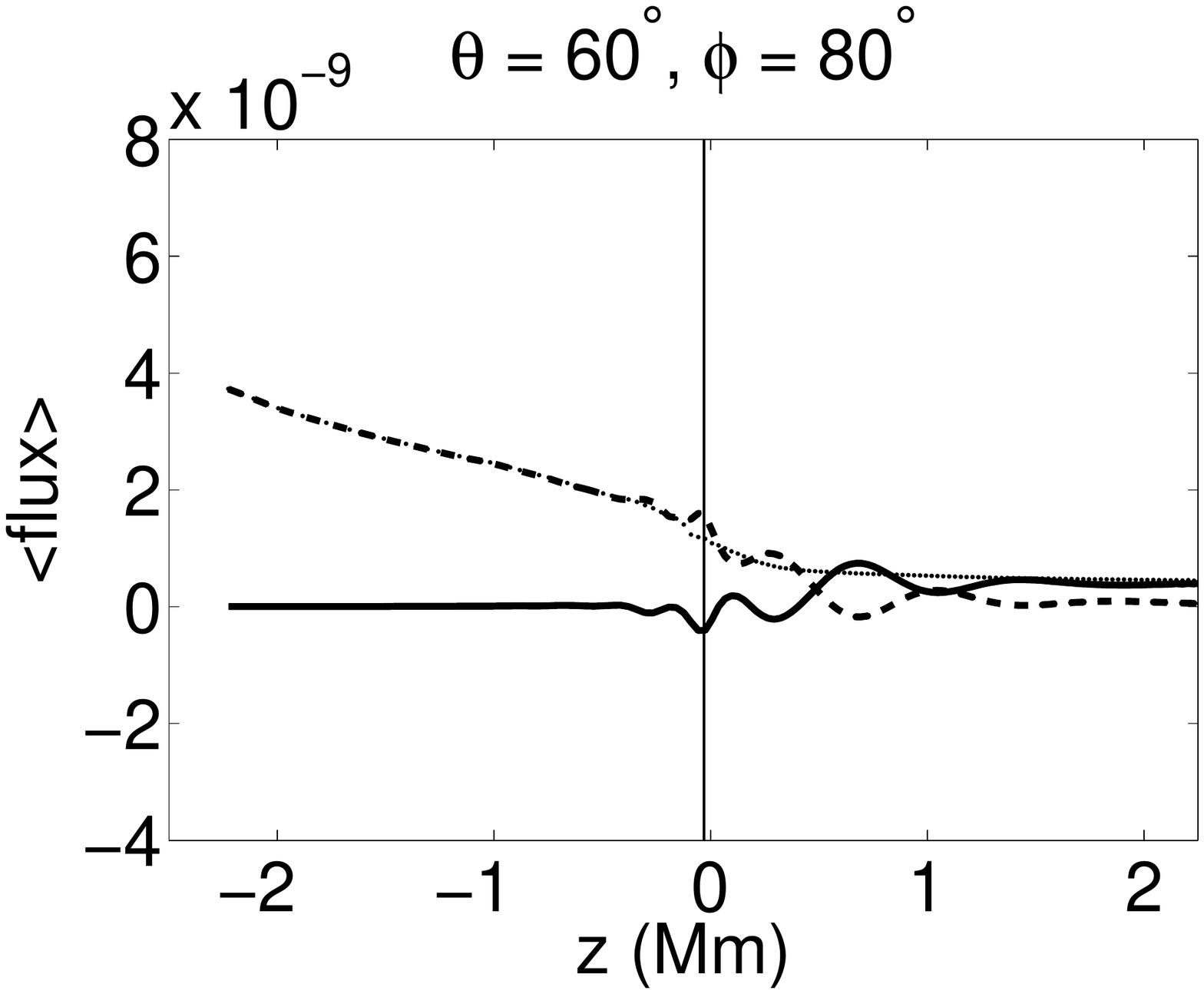}
\end{minipage}\hspace{-1pc}%
\caption{\label{fig:fluxes} Time-averaged magnetic (bold sold line), acoustic (dashed line) and total (dotted line) fluxes (in non-dimensional units) calculated as a function of height. The solid vertical line represents the $c_A \approx c_S$ height.}
\end{figure}

The results of the projected velocities derived from simulations where $\phi$ is fixed at $80^\circ$ and $\theta$ varies from $0^\circ$ to $30^\circ$ to $60^\circ$ are shown in Figure~\ref{fig:proj}. The inclination of the ridges in these projections indicates the wave propagation speed: the more inclined the ridges, the lower the propagation speed and vice versa. The presence of the (primarily acoustic) slow mode ($V_{long}$) is clearly visible above the $c_A \approx c_S$ level (solid line) in all three cases (with the reduced $\omega_{ac}$ resulting in a significant amount of acoustic waves propagating along the field both above and below the equipartition height), while the rapidly propagating Alfv\'en mode ($V_{perp}$) only appears when the magnetic field is sufficiently inclined and oriented out of the plane, i.e, $\theta=30^\circ, \phi=80^\circ$ and $\theta=60^\circ, \phi=80^\circ$, with the latter configuration appearing to be the more efficient in producing Alfv\'en waves. For these two cases, a faint presence of the magnetically dominated fast mode ($V_{trans}$) above the reflection height (dotted line) can still be made out. This is a result of our plane-parallel driver, which excites all wavenumbers, leading to a proportion of fast modes being transmitted through to the PML, rather than being reflected at $c_A Ê\approx Ê\omega/k_h$. In the $\theta=60^\circ, \phi=80^\circ$ projection, there appears to be a wavefront with an opposite inclination of ridges close to the top boundary of the domain. This could either be an artefact of the colour scheme/scaling, or a numerical artefact (i.e, reflection) from the upper boundary condition of the simulations. While we do not observe any reflection from the upper boundary in the corresponding acoustic flux (see Figure~\ref{fig:fluxes}), given that the wavefornt appears to arrive at the upper boundary prior to the arrival of the slow waves, it could also be possible that the signature is a yet unmodelled product of the fast-Alfven conversion process. This is something which we hope investigate in a future work.

Figure~\ref{fig:fluxes} shows the vertical component of the averaged fluxes as a function of height. We observe that the flux variations are strongest near the conversion layer (solid vertical line), with the magnetic flux exceeding the acoustic flux when $\theta=60^\circ, \phi=80^\circ$ for $z > 1$ Mm. These results are in good agreement with previous numerical simulations of fast-Alfv\'en mode conversion using homogenous inclined magnetic fields (e.g., see Figure 3 from \cite{kc2011}). 

\begin{figure}[ht]
\begin{minipage}{13.5pc}
\includegraphics[width=13pc,trim=0cm 6cm 1cm 8cm, clip=true]{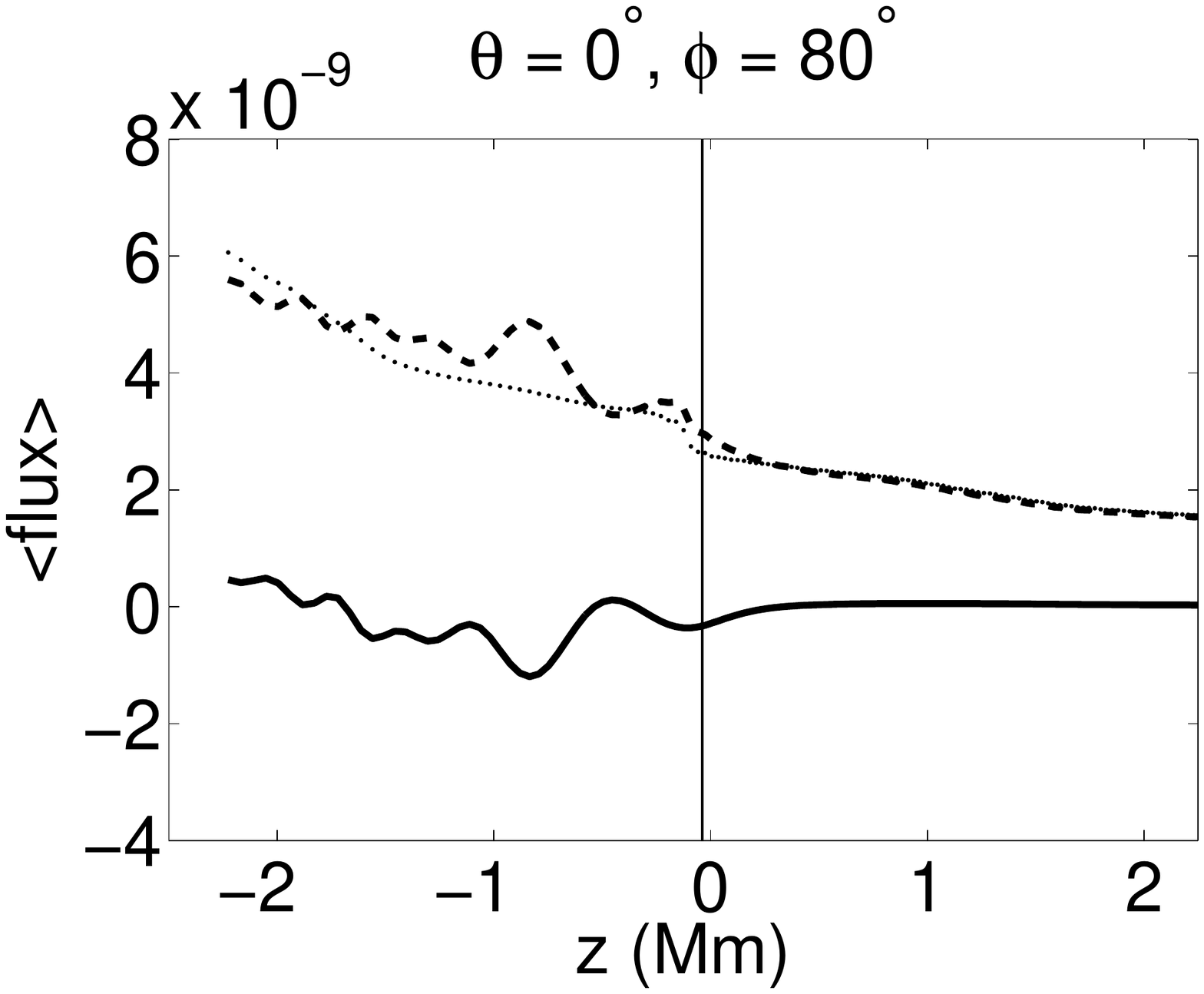}
\end{minipage} \hspace{-1pc}%
\begin{minipage}{13.5pc}
\includegraphics[width=13pc,trim=0cm 6cm 1cm 8cm, clip=true]{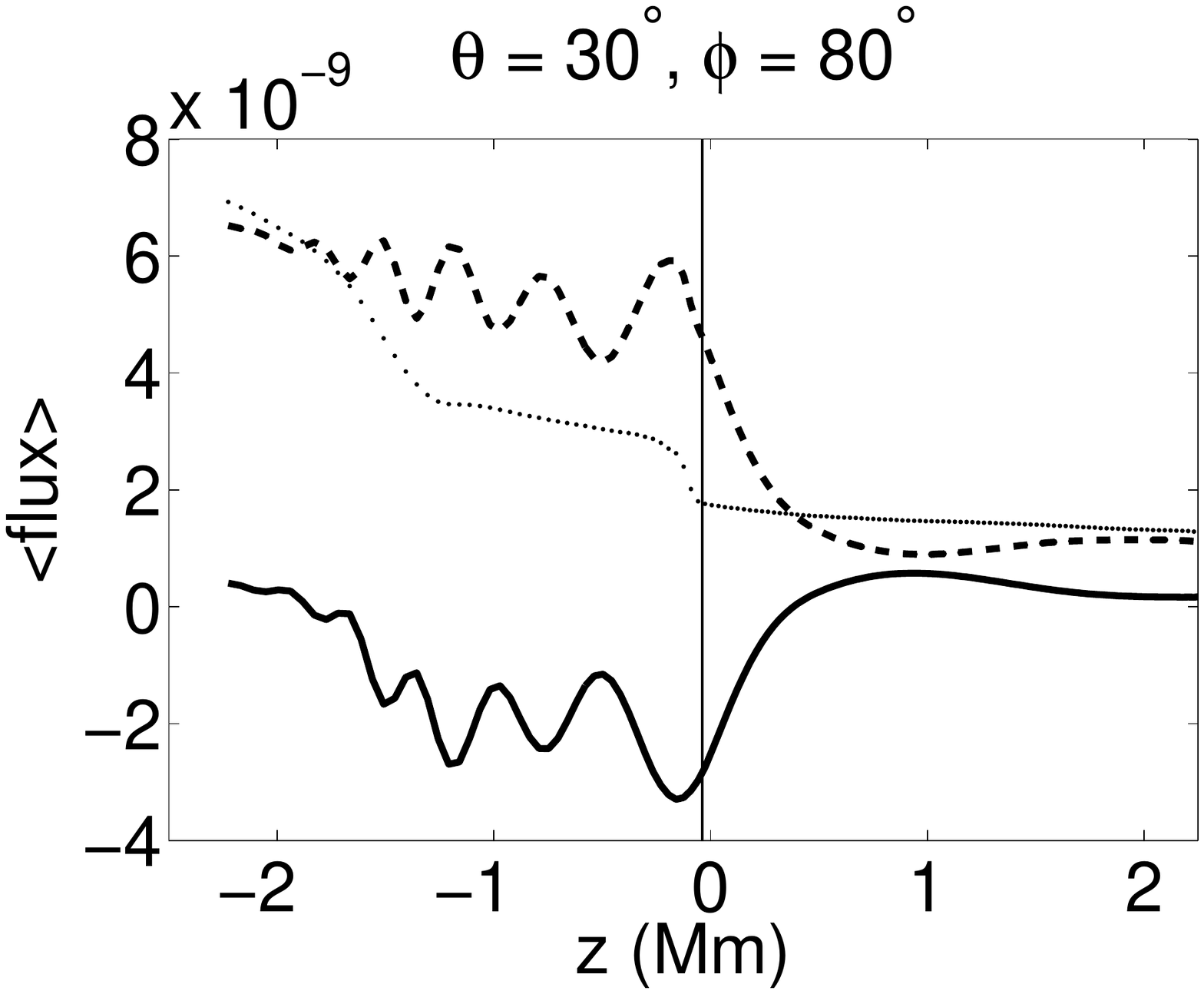}
\end{minipage} \hspace{-1pc}%
\begin{minipage}{13.5pc}
\includegraphics[width=13pc,trim=0cm 6cm 1cm 8cm, clip=true]{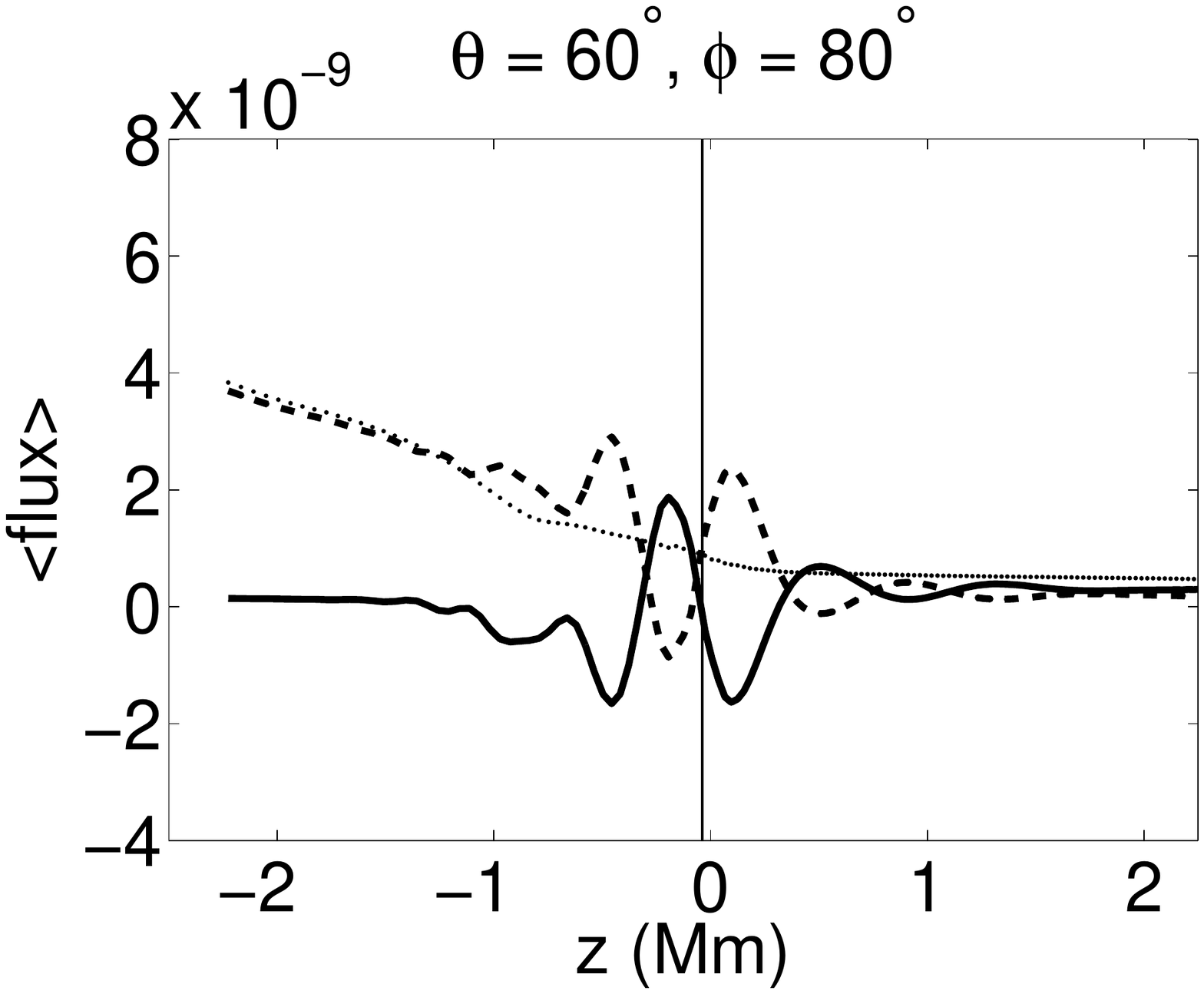}
\end{minipage} 
\caption{\label{fig:fluxes_oldred} Time-averaged magnetic (bold sold line), acoustic (dashed line) and total (dotted line) fluxes (in non-dimensional units) calculated as a function of height for simulations where a Lorentz Force limiter is used with $c_A$ and $c_F$ capped as shown in Figure~\ref{fig:speeds} b). The solid vertical line represents the $c_A \approx c_S$ height.}
\end{figure}

For comparison purposes, we also conducted simulations where, instead of modifying $\rho_0(z)$ and $g_0(z)$ above the surface, we employ a Lorentz Force limiter (with $c_A$ and $c_F$ capped at $\approx 80$ km/s, i.e, as shown in in Figure~\ref{fig:speeds} b)). The resulting averaged fluxes are shown in Figure~\ref{fig:fluxes_oldred}. While differences in the magnitude and height variations of the acoustic fluxes between these results, and those contained in Figure~\ref{fig:fluxes}, can be explained by the change in $\omega_{ac}$, the differences in the magnetic fluxes, particularly when considering the $\theta=60^\circ, \phi=80^\circ$ cases, are almost entirely due to the Lorentz Force limiter. With the Lorentz Force limiter in place, the magnetic flux above $z>1$ Mm appears to just be able to creep above the acoustic flux for a couple of hundred kilometres, before being completely damped prior to reaching PML. We observed this phenomenon regardless of the value of the $c_A$ cap that was used with the Lorentz Force limiter. As expected, the resulting velocity projections for these cases (figures not included) also confirmed the absence of any significant Alfv\'en modes above $c_A \approx c_S$. 

\section{Summary}
Understanding the physics of propagating waves within regions of strong magnetic fields, of which fast-Alfv\'en wave mode conversion has recently been shown to be a critical component, is essential for helioseismic studies of sunspots and active regions. We used the 3-D linear MHD solver SPARC to simulate this process in a convectively stabilised solar model (CSM\_B with empirically modified $\rho_0(z)$ and $g_0(z)$ profiles in the upper layers to ensure a reasonable $\Delta t$) permeated by homogenous inclined magnetic fields. We found that employing a traditional Lorentz Force limiter to artificially cap $c_A$ above the surface tends to inhibit the fast-Alfv\'en mode conversion process by significantly damping the magnetic flux above the surface. 

The next steps in our forward modelling process will include the introduction of random stochastic sources and more realistic (i.e., sunspot-like) background atmospheres, in order to simulate artificial helioseismology data sets. With the aid of local helioseismic diagnostic tools, such as time-distance helioseismology and helioseismic holography, we will then be able to attempt to quantify the effects of fast-Alfv\'en mode conversion on the wave travel times. 

\ack

This work was supported by an award under the Merit Allocation Scheme on the NCI National Facility at the ANU. 
\\


\providecommand{\newblock}{}

\end{document}